\documentclass[aps,amsfonts,nofootinbib,superscriptaddress,preprint]{revtex4-1}
\usepackage{comment}
\usepackage[normalem]{ulem} 
\makeatletter
\usepackage{amsmath}
\usepackage{amssymb}
\usepackage{appendix}
\usepackage{amsbsy}
\usepackage{bm}
\usepackage{epsfig} 
\usepackage{color}
\usepackage{slashed}
\usepackage{soul}
\usepackage{comment}
\usepackage[normalem]{ulem} 
\makeatletter
\usepackage{amsmath}
\usepackage{amssymb}
\usepackage{appendix}
\usepackage{amsbsy}
\usepackage{bm}
\usepackage{epsfig} 
\usepackage{color}
\usepackage{slashed}
\usepackage{soul}
\newcommand{\ee}{\end{equation}}
\newcommand{\bb}{\begin{equation}}
\newcommand{\eqb}{\begin{eqnarray}}

\newcommand{\eqf}{\end{eqnarray}}

\def\nablavec{\mbox{\boldmath$\nabla$}}

\DeclareMathOperator{\Tr}{Tr}
\usepackage[T1]{fontenc}

\begin{document}
\title{ Non-Perturbative Aspects of  Spontaneous Symmetry Breaking}
\author{J.~Gamboa}
\email{jorge.gamboa@usach.cl}
\affiliation{Departamento de  F\'{\i}sica, Universidad de  Santiago de
 Chile, Casilla 307, Santiago, Chile}
\author{J. L\'opez-Sarri\'on}
\email{jujlopezsa@unal.edu.co}
\affiliation{Departamento de F\'{\i}sica Te\'orica, At\'omica y \'Optica, Universidad de Valladolid,
47011 Valladolid, Spain
\\and \\
Departamento de F\'{\i}sica, Universidad Nacional de Colombia,
111321 Bogot\'a, Colombia}

\begin{abstract}
Spontaneous symmetry breaking is studied in the ultralocal limit of a scalar quantum field theory, that is when $E\approx m$ (or infrared limit). In this infrared approximation the theory $\varphi^4$ is formally two-dimensional and its Euclidean solutions are instantons. For BPST-like solutions with $A^a_\mu = A^a_\mu (x^2)$,  the map between 
$\varphi^4$ in two dimensions and Self-dual Yang-Mills theory is carefully discussed.
\end{abstract}
\maketitle
\section{Introduction}

In recent years an extensive discussion of how to implement the interaction between visible and dark matter \cite{1,2,3,4,5,6,7,8} has taken place and in this direction the role played by the Higgs portal has been fundamental \cite{portal,portal1,portal2,portal3,portal4,portal5,portal6,portal7}. The Higgs Portal is one of the possible prescriptions to implement the interactions between visible and dark matter in the extensions of the standard model. Among its important properties is the renormalizability and, of course, the simplicity to generate masses following the same ideas as in the conventional Higgs mechanism.

In the non-perturbative sector there are many important works since the discovery of the instanton solution \cite{poly1}, quantum corrections  \cite{hooft,hooft1,fidel}, sphaleron solution \cite{manton}, $\theta$-vacuum \cite{dashen,jackiw1,jackiw2} and so on \cite{actor}.

However, there are also other non-perturbative aspects of the standard model associated with infrared problems  where an effort to understand this regimen is clearly necessary. From the physical point of view there are at least two reasons for us to investigate in this point;  the first, is because the interaction between visible and dark matter is very weak and non-perturbative considerations on the Higgs portal can be relevant.  The second one,  is because an analysis including local and global aspects of the Higgs portal is not only useful, but is also important for the analysis of spontaneous symmetry breaking and the role of vacuum beyond perturbation  theory.

In this context we would like to explain the problem we will solve in this paper; the vacuum is understood in quantum field theory as the  state of lowest energy from which all other many particles states are constructed. From this perspective and depending on the phenomenon, the vacuum can be unique and be the starting point for a perturbative treatment of a particular quantum field theory. 

However, if there is spontaneous symmetry breaking, it could happen that perturbation theory is not applicable and the question is, how do we proceed?. 

In order to explain the situation, 
let us start considering the Lagrangian 
\bb 
  {\cal L} = \frac{1}{2} (\partial \varphi)^2 +V (\varphi), \label{2}
  \ee
  where 
  \bb 
V (\varphi) = -\frac{1}{2}\mu^2 + \frac{\lambda}{4} \varphi^4. \label{2x}
\ee   
This potential has two minima  
\bb
\varphi_0=v=\pm \sqrt{\frac{\mu^2}{\lambda}},  \label{1x}
\ee
and an unstable extreme in $\varphi_0 =0$.

However in the presence of spontaneous symmetry breaking the approach (perturbative) is as follows; since $v =\langle 0|\varphi|0\rangle \neq 0$, we shift $\varphi$ to
\bb
\varphi= v+ {\bar \varphi}, 
\ee
where ${\bar \varphi}$ is a fluctuation around the vacuum $v$. 

If the fluctuations are small, we can make perturbation theory as in the standard model, but if the minima are very deep, the low energy excitations will be \lq \lq trapped\rq \rq ~at the bottom of the wells. For the latter case the choice $+v$ or $-v$ might produce physically non-equivalent results unless an instantons tunneling between vacua takes place \cite{polyakov,coleman1,coleman2,coleman3,marino,manu,fradkin}.

At first glance the instantons tunneling problem would be difficult to study unless we consider field theory in $ 2d $ dimensions. However there is a heuristic way of looking at this problem which is by considering the infrared  limit of a field theory \cite{klauder}.  This tunneling cannot occur because the quantization volume is infinite and the tunneling probability for infinite volume is zero. However, this is a subtle point because we should first determine the scale that infrared phenemomena occur and then see that $\frac{\ell}{L }\ll 1$ (see below).

Roughly speaking the infrared limit implies that in the dispersion relation $E^2 ={\bf p}^2 +m^2$ we can assume that $E^2 \approx m^2$, and instead of the Lagrangian 
\bb
{\cal L}= \frac{1}{2}(\partial \varphi)^2 -\frac{1}{2} m^2\varphi^2 +\cdots,  \label{eff1}
\ee
we can write, 
\bb
{\cal L}_0 = \frac{1}{2}{\dot \varphi}^2 -\frac{1}{2} m^2\varphi^2 +\cdots \label{eff2}
\ee
in other words we neglect the spatial derivatives. 

However this abrupt transition comes from the following effective Lagrangian
\bb
{\cal L}_{eff} = {\cal L}_0+ \sum_{n=0}^\infty c_n\left(\frac{\ell}{L}\right)^{2n} \left(\nabla \varphi\right)^{2n}, 
\ee
where $c_n$ are coefficients that must be calculated using the symmetries of the effective field theory.

We will point out below that the infrared limit of a scalar field theory maps exactly to a Yang-Mills theory with self-dual solutions. Showing details and physical assumptions is one of the goals of this paper.

In the second part of our work, we do a global analysis for a Higgs portal and discuss the implications that this analysis has in the context explained above. So we show in the following sections how in the non-perturbative sector of the Higgs portal there is also tunneling of instantons.

The paper is organized as follows: in the next section we will explain how instantons emerge in the infrared regime, in section III and IV we will generalize our results for a Higgs Portal and analyze the structure of vacuum, in section V we will discuss  the role played by the instantons in the Higgs portal description and the section VI contains the conclusions.

\section{ Instantons and spontaneous symmetry breaking}

\subsection{Infrared approximation}

As we explained slightly above, the infrared approximate consists roughly speaking throwing out the spatial derivatives of the Lagrangian
\eqb
{\cal L}  &=& \frac{1}{2} (\partial \varphi)^2 + U(\varphi) \nonumber 
\\
&=& {\dot \varphi}^2 + \frac{1}{2} (\nablavec \varphi)^2 + U(\varphi).  \label{0x}
\eqf

In general, most of the fluctuations will   not be able to    yield     vacuum tunneling; However in the infrared approximation, that we are considering in this paper some particular ones will have a non negligible probability of this effect. In particular for those fluctuations which change rapidly  in a  spatial volume which is  large compared with the volume accesible for the fluctuation but small  enough compared with the typical   spatial volume  where the  variation of the fluctuation is sensitive,  then the  instanton mechanism will be possible.

 In order to explain last ideas,  let us consider the fluctuation  $\varphi(t,\vec{x})$.  The typical time and length scales of the fluctuation  can be estimated by,

\begin{eqnarray}
\frac{1}{T}&\sim& \left|\frac{\partial_t\varphi}{\varphi}\right|,\\ 
\frac{1}{L}&\sim&\frac{\vert \vec\nabla\varphi\vert}{|\varphi|}.
\end{eqnarray}
Hence  fast fluctuations correspond to the condition \footnote{If we take $\varphi \sim e^ {i k.x}$ this approximation is $E \gg |{\bf p}|$.},
\begin{equation}
T\ll L\,,
\end{equation}
which is equivalent to neglect spatial derivatives compared to time derivatives of $\varphi$ in the lagrangian density. 

The next step in our argument is to note that the equation of motion for the theory $\varphi^4$ in this infrared approximation is
\begin{equation}
{\ddot  \varphi} = \mu^2 \varphi - \lambda  \varphi^3,  \label{scale}
\end{equation}
In order to make dimensionless the field $\varphi$ we multiply (\ref{scale})  by the length scale $\ell$ and obtain
\begin{equation}
{\ddot {\bar \varphi}} = \mu^2 {\bar \varphi} - {\bar \lambda} {\bar \varphi}^3, \label{eqq}
\end{equation}
where ${\bar \varphi}= \ell \varphi$ and the effective coupling constant is 
\begin{equation}
{\bar \lambda} = \frac{\lambda}{\ell^2}. 
\end{equation}

The solution of (\ref{eqq}) is 
\begin{equation}
{\bar \varphi}(t) = \frac{\mu ~\ell}{\sqrt{\lambda}} \tanh \left[ \frac{\mu}{\sqrt{2}} (t -t_0)\right]. \label{arg}
\end{equation}

Note that now the wave amplitude is dimensionless and the infrared approximation implies that
$\mu \ell \ll 1$ and the infrared scale length is
\begin{equation}
\ell \ll \frac{1}{\mu}\sim  T\ll L\,, \label{sca}
\end{equation}
where we  have taken into account that the characteristic time scale $T$ of the solution (\ref{arg})  is $T=\sqrt{2}/\mu$ which  is much smaller than the characteristic length scale $L$ of the fluctuation.

   With these results in mind, we can see that the tunneling probability es 
   \begin{equation}
   P = e^{-2 \int_{-a}^a d{\bar \varphi}\sqrt{2U({\bar \varphi)}}} = e^{- \frac{8}{3} \frac{\mu^3 \ell^3}{\lambda}},  \label{prob}   \end{equation} 
 where we have used $a= \frac{\mu \ell}{\sqrt{\lambda}}$.  
 
 We should note that the volume $v=\ell^3$, which appears in the formula for the probability of tunneling,  is a consequence of the range of energies (infrared) that we are considering. Although   this volume may be large, it is finite and much smaller than the characteristic volume $V=L^3$, which can be arbitrarily  large,  and therefore, the probability of tunneling cannot be neglected (in this approximation).

Using the above arguments we see that 
\bb
{\cal L} = \frac{1}{2} {\dot \varphi }^2 + \frac{\bar \lambda}{4}\left({\bar \varphi}^2 - \frac{\mu^2}{\bar \lambda}\right)^2, \label{4x}
\ee 
and this last Lagrangian describes the infrared sector and  to throwing away the spatial derivatives of the Lagrangian is equivalent to the ultralocal limit  (infrared) \cite{klauder}.  This infrared limit has been used also in different contexts in quantum gravity \cite{pilati1,pilati2,pilati3,null1,null2,null3,isham,anderson}.

As a final comment to this subsection, it is interesting to note that the condition (\ref{sca}) of course implies 
\bb
\frac{\ell}{L} \ll 1, \label{condicion1}
\ee 
which is --in an explicit calculation-- the natural expansion parameter.

\subsection{Scalar and Yang-Mills theories} 

In this subsection we will give direct and simple arguments to show the mapping between scalar and $SU(2)$ self-dual Yang-Mills theory. 

First let us  write the Lagrangian density (\ref{4x}) 
in terms of dimensionless variables by the rescaling 
\[ 
{\bar \varphi} \to \frac{ \sqrt{\lambda}}{\mu} {\bar \varphi}, ~~~~~t \to \mu~ t 
\] 
 and the action becomes
\bb
S_S = \frac{1}{\lambda} \int d{\bar t} \left[\frac{1}{2} {\dot {\bar \varphi}}^2+ \frac{1}{4} \left({\bar \varphi}^2 - 1 \right)^2 \right]. \label{ac1}
\ee

Now we could compare this action with the Yang-Mills one taking into account the following:
\begin{itemize} 
\item Since the Yang-Mills description we are looking must contain instanton solutions and the self-duality condition
\bb
F_{\mu \nu} = {\tilde F}_{\mu \nu} \label{selfdua}
\ee
must be satisfied.
\item  If we are able to find a potential that satisfies (\ref{selfdua}), then we not only solve the condition (\ref{selfdua}) but we automatically have a solution of the Yang-Mills equations. Since in the Euclidean space $SO(4) \simeq SU(2) \times SU(2)$, the Yang-Mills potentials are a possible representation of $ (\frac{1}{2}, \frac{ 1}{2}) $ of $SU (2) \times SU(2)$. 

The tensor $F_{\mu \nu}$  belong to a reducible representation $(1,0) \oplus (0,1)$ so that the conditions of self and anti-self dualities can be represented by the quantities 
\eqb 
F^a_L &=& \eta^a_{\mu \nu} \left( F_{\mu \nu} + {\tilde F}_{\mu \nu}\right), 
\\
F^a_R &=& {\bar \eta}^a_{\mu \nu} \left( F_{\mu \nu} -{\tilde F}_{\mu \nu}\right), 
\eqf
where $\eta$ and ${\bar \eta}$ are the `t Hooft symbols for $SU(2)$  which have the following properties: 
\eqb 
\eta_{a \mu \nu} &=&  \frac{1}{2} \epsilon_{\mu \nu \rho \sigma} \eta_{a \rho \sigma}, \nonumber
\\ 
{\bar \eta}_{a \mu \nu} &=& -\frac{1}{2} \epsilon_{\mu \nu \rho \sigma} {\bar \eta}_{a \rho \sigma} \nonumber 
\\
{\bar \eta}^a_{\mu \nu} (x)  &=&\epsilon^a_{\mu \nu 4} -\delta^a_\mu \delta_{\nu 4}+ \delta^a_\nu \delta_{\mu 4}, \nonumber
\\
\epsilon_{abc} \eta_{b \mu \nu}  \eta_{c \rho \sigma}&=& \delta_{\mu \rho} + \delta_{\nu \sigma} \eta_{\nu \sigma} \eta_{a \nu \rho} -\delta_{\nu \rho} 
\eta_{a \mu \sigma}, \nonumber
\eqf
among other \cite{van}.
\end{itemize} 

Condition (\ref{selfdua}) is a well-known result \cite{poly1} but finding a potential that satisfies (\ref{selfdua}) is a highly non-trivial problem that it was solved  by Diakonov in \cite{diakonov}. 

The potential is 
\bb
A^a_\mu = {\bar \eta}^a_{\mu \nu} x_\nu \frac{[1+ \phi (x^2)]}{x^2}. \label{pot}
\ee 
 where $\phi$ a scalar field.
 
 This ansatz is generally only shown in review papers but the important steps in between are never explained, here we will give a derivation of these results and show how the BPST (and Diakonov) solution relates to each other with 2D instantons. 

The first step to justify Diakonov's ansatz is to write the $SU(2)$ Yang-Mills potential as 
\bb
A^a_\mu = \eta^a_\mu (x) g(x^2). \label{xx1}
\ee
where $ \eta^a_\mu (x)$ is a function that we will determine below using the t'Hooft symbols and $g(x^2)$ is a function to be determined.

The function $\eta^a_\mu$ is obtained as follow: firstly we take the t'Hooft symbols definition
\[
{\bar \eta}^a_{\mu \nu} (x)  = \epsilon^a_{\mu \nu 4} -\delta^a_\mu \delta_{\nu 4}+ \delta^a_\nu \delta_{\mu 4}
\]
and multiplying by $x_\nu$ we obtain 
\bb
\eta^a_\mu (x) = x^a \delta_{\mu 4} - x^4 \delta_{\mu 4} + \epsilon^{a jk} \delta_{\mu j} x_k. \label{xx3}
\ee

By using (\ref{xx1}) and (\ref{xx3}) the tensor $F^a_{\mu \nu}$ becomes 
\bb
F^a_{\mu \nu} = g(x^2)  \left( \partial_\mu \eta^a_\nu -\partial_\nu \eta^a_\mu \right) + 2 g'(x^2) \left( x_\mu \eta^a_\nu - x_\nu \eta^a_\mu\right) + g(x^2) \epsilon^{abc} \eta^b_\mu \eta^c_\nu. \label{xx4}
\ee

Here $g' (x^2) = \frac{d}{dx^2} g(x^2)$ and the Yang-Mills equations become 
\eqb
\partial_\mu F^a_{\mu \nu} (x^2) &+& \epsilon^{abc} A^b _\mu F^c_{\mu \nu} =  \nonumber
\\
&=&\left( 4 x^2 g''(x^2) + 12 g'(x^2) - 2 x^2 g^3 (x^2) \right) \eta^a_\nu + 3g^2 (x^2) \epsilon^{abc} \eta^b_\mu \partial_\mu \eta^c_\nu =0. \nonumber 
\\ 
\label{xx5}
\eqf

If we choose all the components of the group different, we obtain the following non-linear differential  equation
\bb
2 x^2 g'' + 6 g' +3g^2 -x^2 g^3=0. \label{xx6}
\ee

In order to solve this eq. let's do the following variable change
\eqb
g(x^2) &=& \frac{h(x^2)}{x^2}   \label{xx7}
\\
x^2 &=& \rho^2 e^t, \label{xx8}
\eqf
where $\rho^2$ is scale parameter that was introduced for dimensional reasons.

Replacing (\ref{xx7}) and (\ref{xx8}) we obtain de following equation
\bb
2 {\ddot h} (t)- 2 h(t) -3h^2(t) - h^3(t)=0.
\ee

These equations can be obtained from the following Lagrangian
\eqb
L &=& {\dot h}^2 + h^2 - h^3 + \frac{1}{4} h^4 \nonumber
\\
&=& {\dot h}^2 + \frac{1}{4} h^2 (h-2)^2,  \label{9}
\eqf
and making the shift
\bb
h \to \varphi +1, \label{xx10}
\ee
we obtain
\bb
L = {\dot \varphi}^2 + \frac{1}{4} (\varphi^2 -1)^2.
\ee

Replacing (\ref{xx7}) and (\ref{xx10}) in (\ref{pot}) we obtain de Diakonov's ansatz but we also obtain that $\varphi$ is a solution of the instantons in $2D$ equations.

Now we replacing (\ref{pot}) in the Yang-Mills action
$$
S_{YM} = -\frac{1}{4g^2} \int d^4x ~\Tr (F_{\mu \nu}^2),
$$ 
we find 
\bb
S_{YM} = \frac{12 \pi^2}{g^2} \int d\tau \left[ \frac{1}{2}\left(\frac{d\phi}{d\tau}\right)^2 + \frac{1}{4} (\phi^2 -1)^2\right]. \label{ac2}
\ee
where $\tau = \ln \frac{x^2}{\rho^2}$. 

Comparing (\ref{ac1}) and  (\ref{ac2}) we see that both actions are equivalent if
\bb
\lambda= \frac{g^2}{12\pi^2}. 
\ee

We finish this section by noting that the $ x ^ 2 $ dependence of the gauge potential corresponds to a set of solutions to the Yang-Mills equations that maps 
exactly to the infrared limit.

\section{Higgs Portal and Vacuum} 

 In this section and the  next we analyze the vacuum structure of the Higgs Portal in order to apply and generalize  the results in previous sections.
 
Let us consider two complex scalar field, we say $\phi_1$ and $ \phi_2$ and the Lagrangian 

\begin{equation} {\cal L} = \frac{1}{2} \vert \partial \phi_1\vert^2 +\frac{1}{2}\vert \partial \phi_2\vert^2 -  \frac{1}{2} \mu^2_1\vert\phi_1\vert^2 -  \frac{1}{2} \mu^2_2\vert \phi_2\vert^2 +\frac{\lambda_1}{4}\vert\phi_1\vert^4 + \frac{\lambda_2}{4}\vert\phi_2\vert^4 +\frac{\gamma}{2} \vert\phi_1\vert^2\vert\phi_2\vert^2\,, \label{energyDensity} 
\end{equation}
 where we have introduced an interaction between the two scalar fields by the coupling $\gamma$. 
 
 For $\gamma=0$ we have two copies of a  Higgs model with symmetry breaking given by a Mexican hat potential. The  classical vacuum  is given for the configurations where the potential  reach a minimum. However the structure of these minimal configurations will depend on the values of the parameters of the potential. Thus here we will study how structure of these minima   change in the parameter space, {\it i.e.}, the coupling $\gamma$, the self-couplings $\lambda_i$ as the couplings $ \mu^2_i$. 
 
 The study of the regions of the parameter space which characterize certain vacuum structure is similar to the analysis of the phase diagram of a thermodynamical system with second order transitions. So, in many situations we will use the terminology of the thermodynamical phenomena in the Ginzburg-Landau theory. Thus, our analysis is similar to that of  a system thermodynamical potential depending on an order parameter is characterized by the couplings $\mu^2$ of a quadratic term and   $\lambda>0$ of a quartic contribution.   The point  where the coefficient $\mu^2$ changes of sign represents the critical point of a second order phase transition so that the ground  state characterized by a vanishing (and symmetric) order parameter, changes to the situation, where the minimal configuration consists of a non- vanishing (symmetrically broken) order parameter. 
 
 We will study the possible diagram of phases in the space of the three parameters $\lambda_i$ and $\gamma$ fixed $\mu^2_i$ to be positive, but we will briefly discuss how this diagram changes when other signs for $\mu^2$ are considered.

To perform  that analysis it is enough to   consider only the minima of the density potential, 
\begin{equation} 
V=- \frac{1}{2}\mu^2_1\vert\phi_1\vert^2 -  \frac{1}{2}\mu^2_2\vert \phi_2\vert^2 +\frac{\lambda_1} {4}\vert\phi_1\vert^4 +\frac{\lambda_2}{4}\vert\phi_2\vert^4 +\frac{\gamma}{2} \vert\phi_1\vert^2\vert\phi_2\vert^2\,,\label{potentialDensity} 
\end{equation}
 where we have neglected the derivative terms in (\ref{energyDensity}).  
  In this section  we will study the global stability of the model together with the phase structure depending on the parameter $\gamma$ characterizing the vacuum states and their   symmetries.

  \subsection{Global stability} 
  
  To study the global stability we will search for   the regions of the parameter space   where  the density potential (\ref{potentialDensity}) is bounded from below. Because this potential is a polynomial function on the fields,  , we must look at large values of $ \vert\phi_1\vert$ and $\vert\phi_2\vert$. In that region the quadratic terms are very small in comparison with the quartic terms. So we must take into account the last three terms of (\ref{potentialDensity}). A sufficient condition for the potential to be bounded is then that for any  
   direction in the plane $ \vert\phi_1\vert-\vert\phi_2\vert$ the  profile of the potential is increasing with the field norm. So, if we take, 
  $$\vert\phi_2\vert=m\vert\phi_1\vert$$ 
  with $m\in[0,\infty)$ being the slope which parametrizes the chosen direction, and  substituting this on the quartic terms, we obtain, 
$$\left(\frac{\lambda_1}{4} + \frac{\gamma}{2} m^2 + \frac{\lambda_2}{4} m^4\right) \vert \phi_1\vert^4\,.$$
   Thus, the density potential will be globally stable if the coefficient in the expression between parenthesis is positive for the whole   range of $m$.
   For the cases of $m=0$ or $m\to\infty$ we see that $\lambda_1$ and $\lambda_2$ must be positive. Furthermore, the positivity of that coefficient  will be ensured if, also,  there are no real and positive zeros for $m^2$.  The zeros are,
$$m^2=\frac{-\gamma\pm\sqrt{\gamma^2-\lambda_1\lambda_2}}{\lambda_1}\,.$$
This happens when   $\gamma$  is positive or $\gamma^2<\lambda_1\lambda_2$.

Summarizing, sufficient  conditions for global stability are, 
\begin{equation}
     \gamma> -\sqrt{\lambda_1\lambda_2}\,,\hspace{1cm}\lambda_1>0,\qquad\lambda_2>0\,.
      \label{globalStability} 
      \end{equation}
      
       Notice that the region $\gamma^2=\lambda_1\lambda_2$ corresponds to a cone surface  in the three dimensional space $\lambda_1-\lambda_2-\gamma$ with the symmetrical axis being the straight line $ \lambda_1=\lambda2$, and $\gamma=0$. 
    
     Note also that these are sufficient but no  necessary conditions. The failure of these conditions can still give rise to a stable model depending on the coefficients $\mu_1^2$ and $\mu_2^2$. However, in the most of our analysis we take positive values of the $\mu^2$ factors, and the above conditions are actually necessary and sufficient.

\subsection{Ground  States Structure}

 In this section  we will look for the homogeneous configurations which minimize    the   energy density (\ref{energyDensity}), or equivalently, the potential density (\ref{potentialDensity}). 
 
 To do this, let us write the complex scalar fields as,
\begin{eqnarray}
\phi_1&=&\rho_1\,e^{i\varphi_1}\,,\label{phi1}\\
\phi_2&=&\rho_2\,e^{i\varphi_2}\,,\label{phi2}
\end{eqnarray}
 where $\rho_1,\rho_2\geq0$ and $\varphi_1,\varphi_2$ are phases. It is evident that the density potential is independent of the phases and they  will not have any role in our discussion (up to the degeneracy of the possible ground states). Then, in order to find local extrema we proceed as usual by derivating with respect coordinates $\rho_1$ and $\rho_2$, equaling them to zero and check that the hessian matrix at those points is positive definite. However, and because the  domain of the variables $\rho_1$ and $\rho_2$ are not open regions we must to take into account also  possible minima at the boundary  ($\rho_1=0$, or $\rho_2=0$).     
 
 Hence, derivating with respect to $\rho_1$ and $\rho_2$ and equaling them to zero gives, 
 \begin{eqnarray}
 \frac{\partial V}{\partial\rho_1}&=&\rho_1\left(-\mu_1^2 +\lambda_1\rho_1^2 + \gamma \rho_2^2\right)=0\,,\label{derivativeRho1}\\
 \frac{\partial V}{\partial\rho_2}&=&\rho_2\left(-\mu_2^2 +\lambda_2\rho_2^2 + \gamma \rho_1^2\right)=0\,,\label{derivativeRho2}
\end{eqnarray}
 And the hessian matrix is,
  \begin{equation}
\mathcal{H}(\rho_1,\rho_2)\equiv\left[\frac{\partial^2 V}{\partial\rho_i\partial\rho_j}\right] =
\left(
\begin{array}{c c}
-\mu_1^2+3\lambda_1\rho_1^2+\gamma\rho_2^2& 2\gamma\rho_1\rho_2\\
2\gamma\rho_1\rho_2 &-\mu_2^2+3\lambda_2\rho_2^2+\gamma\rho_1^2
\end{array}\right)\,.
\label{hessian}
\end{equation}

From these last expressions it is 
   clear the following facts:
   \begin{enumerate}
   \item With positive $\mu^2$'s factors there will never be symmetric solutions $\rho_1=\rho_2=0$ due to ${\mathcal H} < 0$.
\item  we will  have solutions of the form,
\begin{equation}
\rho_1^{\prime2}=2\frac{\mu_1^2}{\lambda_1}\,,\hspace{.5cm}\textrm{and,}\,\hspace{0.5 cm}\rho_2^{\prime2}=0\,,
\label{rhoPrime}
\end{equation}
with positive definite hessian if,
\begin{equation}
\gamma>\lambda_1  \frac{\mu_2^2}{\mu_1^2}\,.
\label{condPrime}
\end{equation}
And the solution,
\begin{equation}
\rho^{\prime\prime2}_1=0\,,\hspace{.5cm}\textrm{and,}\hspace{0.5 cm}\rho_2^{\prime\prime2}=2\frac{\mu_2^2}{\lambda_2}\,,
\label{rhoPrimePrime}
\end{equation}
with positive definite hessian if,
\begin{equation}
\gamma>\lambda_2  \frac{\mu_1^2}{\mu_2^2}\,.
\label{condPrimePrime}
\end{equation}

\item If $\rho_1$ and $\rho_2$ are different from zero. Then eqns (\ref{derivativeRho1}) and (\ref{derivativeRho2}) can be written in a matrix form,
\begin{equation}
\left(\begin{array}{c c}
     \lambda_1&\gamma\\ 
     \gamma&\lambda_2 
\end{array}\right)
\left(\begin{array}{c} \rho_1^2\\ \rho_2^2 \end{array}\right) =\left(
\begin{array}{c}
2\mu^2_1\\
2\mu_2^2 
     \end{array}\right)\,. 
     \label{derivativeMatrix}
    \end{equation}
 And hence if the $2\times2$ matrix is regular, then, we solve to find,
\begin{eqnarray}
\rho_1^2&=&\frac{\lambda_2\mu^2_1 -\gamma\mu_2^2}{\lambda_1\lambda_2-\gamma^2}\,,\label{solRho1}\\
\rho_2^2&=&\frac{\lambda_1\mu^2_2 -\gamma\mu_1^2}{\lambda_1\lambda_2-\gamma^2}\,.\label{solRho2}
\end{eqnarray}
where we have to  emphasize that the right hand sides must be positive  to be acceptable solutions. Furthermore, on solutions of this kind the hessian matrix  is reduced to,
\begin{equation}\mathcal{H}=\left(\begin{array}{c c}
2\lambda_1\rho_1^2& 2\gamma\rho_1\rho_2\\
2\gamma\rho_1\rho_2&2\lambda_2\rho_2^2
\end{array}\right)\,,
\label{hessian0}
\end{equation}
whose trace is always positive definite but  its determinant,
$$\textrm{det}\mathcal{H} = 4\left(\lambda_1\lambda_2 -\gamma^2\right)\rho_1^2\rho_2^2\,,$$
is positive only when $\lambda_1\lambda_2>\gamma^2$

 Hence, positivity of $\rho_1^2$, $\rho_2^2$ and the hessian $\mathcal{H}$  occurs only if,
\begin{equation}
\gamma\leq \lambda_1\frac{\mu_2^2}{\mu_1^2}\,,\hspace{0.5cm}\textrm{and,}\hspace{0.5cm}\gamma\leq  \lambda_2\frac{\mu_1^2}{\mu_1^2}\,.
\label{cond00}
\end{equation}

\item Yet another different  kind of solution happens when,
\begin{equation}
\gamma=\sqrt{\lambda_1\lambda_2}\,,\hspace{0.5cm}\textrm{and,}\hspace{0.5cm}\frac{\lambda_1}{\lambda_2}=\frac{\mu_1^4}{\mu_2^4}\,,
\label{degeneracy}
\end{equation}
 and the values of the field modulus  are not completely determined, but they    lie  on the elliptic curve,
 \begin{equation} 
 \frac{\rho_1}{\lambda_1\mu_1^2}+\frac{\rho_2}{\lambda_2\mu_2^2}= \frac{2}{\lambda_1\lambda_2}\,, 
 \label{ellisoipd} 
 \end{equation}
     
 \end{enumerate}
 Thus, we have different possibilities which define different  minima depending on the region of the parameters space. This will be analyzed  in  next subsections.

\subsection{ Phase I}
\begin{figure} 
   \centering \includegraphics[height=6.5cm]{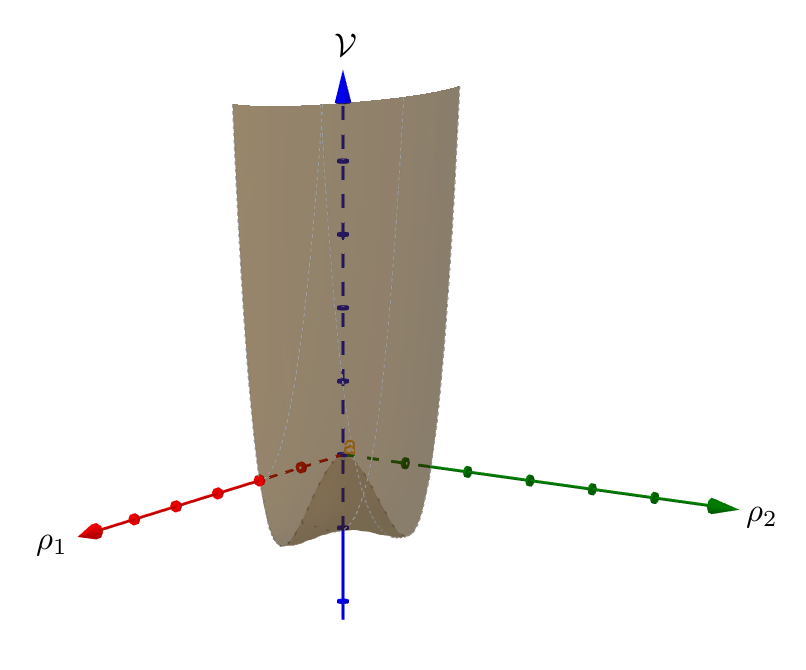}
    \caption[Profile I]{Profile of the potential in phase I}
\label{potential1} 
    \end{figure}

If we restrict to  the region,
\begin{equation}
\gamma>\sqrt{\lambda_1\lambda_2}\,,\hspace{2cm} \frac{\lambda_1}{\lambda_2}=\frac{\mu_1^4}{\mu_2^4}\,,
\label{phaseI}
\end{equation}
which corresponds  to a vertical  half-plane over the cone $\gamma^2=\lambda_1\lambda_2$, as can be seen in (\ref{potential1}), we obtain two equivalent minima at,
$$\rho_1=\sqrt{\frac{\mu_1^2}{\lambda_1}}\,,\hspace{1cm}\rho_2=0\,,$$ 
and,
$$\rho_1=0\,,\hspace{1cm}\rho_2=\sqrt{\frac{\mu_2^2}{\lambda_2}}\,,$$

with potential depth,
\begin{equation}
V_I= -\frac{\mu_1^4}{\lambda_1}= -\frac{\mu_2^4}{\lambda_2}\,,
\label{potentialDepthI}
\end{equation}
and with a minimal  potential barrier given by,
\begin{equation}
\Delta V=\frac{\gamma-\sqrt{\lambda_1\lambda_2}}{\gamma+\sqrt{\lambda_1\lambda_2}}\frac{\mu_1^2\mu_2^2}{\sqrt{\lambda_1\lambda_2}}\,.
\label{barrier}
\end{equation}
This value corresponds to a saddle point placed at,
\begin{eqnarray}
\rho_1^2&=&\frac{ 2\mu_1\mu_2}{\sqrt{\lambda_1\lambda_2}+\gamma}\left(\frac{\lambda_2}{\lambda_1}\right)^{\frac{1}{4}}\,,\label{solRho10}\\
\rho_2^2&=&\frac{2\mu_1\mu_2}{\sqrt{\lambda_1\lambda_2}+\gamma}\left(\frac{\lambda_1}{\lambda_2}\right)^{\frac{1}{4}}\,,\label{solRho20}
\end{eqnarray}

 The profile of the potential can be viewed in figure (\ref{potential1}).

\subsection{Phase II}

  The next situation corresponds to the existence of two minima, when,
  \begin{equation}
  \gamma\geq \lambda_1\frac{\mu_2^2}{\mu_1^2}\,,\hspace{1cm}\gamma\geq\lambda_2\frac{\mu_1^2}{\mu_2^2}\,,
  \label{phaseII}
  \end{equation}
  excluding the half-plane of the phase I. This is the region over the planes $\gamma=\lambda_1\frac{\mu_2^2}{\mu_1^2}$ and $\gamma=\lambda_2\frac{\mu_1^2}{\mu_2^2}$, over the cone $\gamma^2=\lambda_1\lambda_2$. The phase I splits the region into two pieces where there exist two vacua, a false  and the true vacuum (deepest).
 
\begin{figure} 
   \centering \includegraphics[height=6.5cm]{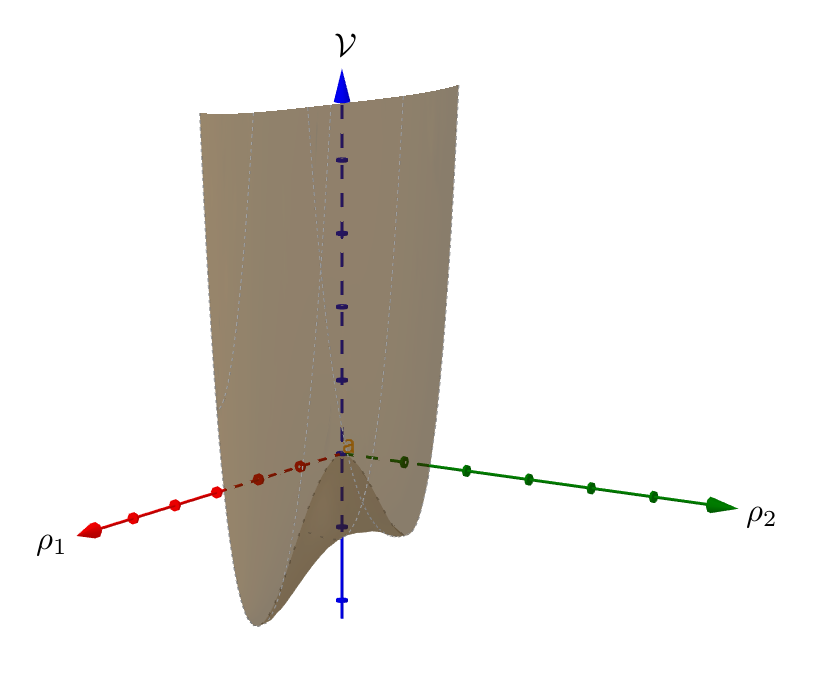}
    \caption[Profile II]{Profile of the potential in phase II}
\label{potential2} 
    \end{figure}

 The left side region in Fig.~(\ref{potential1}) denoted as II,  has the dominant vacuum,

\begin{equation}
\rho_1=\sqrt{\frac{2\mu_1}{\lambda_1}}\,,\hspace{1cm}\rho_2=0\,,
  \label{dominant}
  \end{equation}
  with potential density,
  $$V_{\textrm{II}}=-\frac{\mu_1^4}{\lambda}\,.$$
Also, this phase presents a false vacuum placed at,\begin{equation}
\rho_1=0\,,\hspace{1cm}\rho_2=\sqrt{\frac{2\mu_2^2}{\lambda_2}}\,,
\label{false}
\end{equation}
with potential energy density,
$$V^\prime_{\textrm{II}}=\frac{\mu_2^4}{\lambda_2}\,.$$
  In the middle there is a saddle point placed at points in (\ref{solRho1}) and (\ref{solRho2}), with potential energy density,
  \begin{equation}
 V=-\frac{\lambda_1\mu_2^4-2\gamma\mu_1^2\mu_2^2+\lambda_2\mu_1^4}{\lambda_1\lambda_2-\gamma^2}\,.
 \label{potentialSaddle}
 \end{equation}
 See Fig.~\ref{potential2} to check the characterization of the minima structure in this phase.

  The  right piece, denoted   by II' in Fig.~(\ref{potential1})  is similar  to the situation just above but swapping indices $1$ and $2$, {\it i.e.},  it is characterized  by  a dominant vacuum given by  (\ref{false}), and  a false vacuum given by (\ref{dominant}).

\subsection{Phases III and IV}
Now we analyze the situation where there occurs    a unique vacuum. This happens in two situations.

\begin{figure} 
   \centering \includegraphics[height=6.5cm]{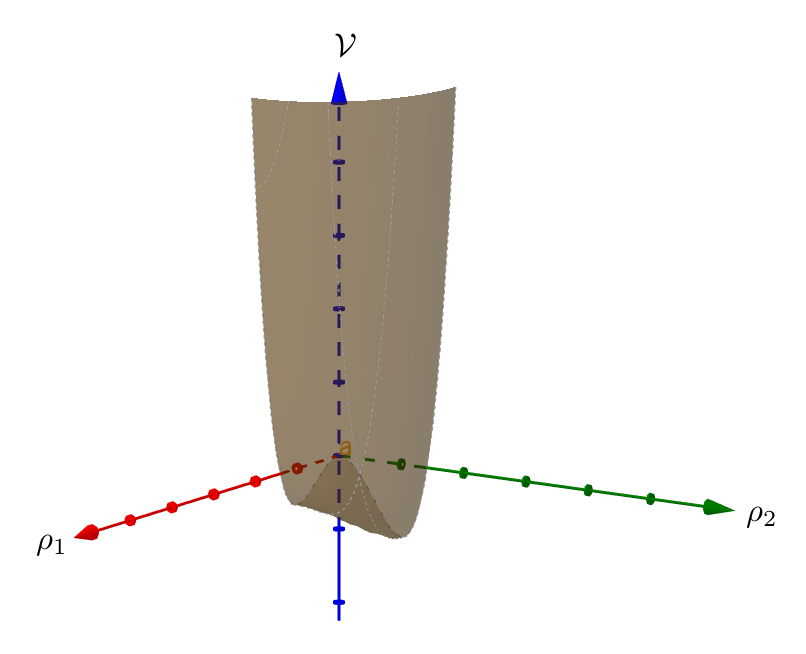}
    \caption[Profile III]{Profile of the potential in phase III}
\label{potential3} 
    \end{figure}

The first situation, phases III and III' in Fig.\ref{potential1}, correspond to the vacuum located at (phase III),
$$\rho_1=0\,,\hspace{1cm}\rho_2=\sqrt{\frac{2\mu_2^2}{\lambda_2}}\,,$$
when 
$$\lambda_2\frac{\mu_1^2}{\mu_1^2}<\gamma<\lambda_1\frac{\mu_2^2}{\mu_1^2}$$
 or (phase III'),
$$\rho_1=\sqrt{\frac{2\mu_2^1}{\lambda_1}}\,,\hspace{1cm}\rho_2=0\,,$$
     when,
  $$\lambda_1\frac{\mu_2^2}{\mu_1^2} <\gamma<\lambda_2\frac{\mu_1^2}{\mu_1^2}$$
     respectively, with minimal potential energies given  in previous subsection. See Fig.\ref{potential3} for the potential profile of this phase.

     The second situation, phase IV in Fig.~\ref{potential1} happens when,
     $$\gamma\leq \lambda_1\frac{\mu_2^2}{\mu_1^2}\,,\hspace{1cm}\gamma\leq\lambda_2\frac{\mu_1^2}{\mu_2^2}\,,$$
     with the unique minimum placed at position given in (\ref{solRho1}) and (\ref{solRho2}) and with potential depth given in (\ref{potentialSaddle}), see Fig.~\ref{potential4}.

  \subsection{Phase V}
   
Finally we obtain an  degenerate vacuum  when,
$$\gamma=\sqrt{\lambda_1\lambda_2}\,,\hspace{1cm}\frac{\lambda_1}{\lambda_2}=\frac{\mu_1^4}{\mu_2^4}\,,$$
which is a straight line corresponding to a generatrix of the cone $\gamma^2=\lambda_1\lambda_2$. The vacuum is degenerate along the elliptic curve (\ref{ellisoipd}) and  its depth potential is,
$$V=-\frac{\mu_1^2\mu_2^2}{\sqrt{\lambda_1\lambda_2}}\,,$$
see Fig.~\ref{potential5}.

\section{Sign changes in $\mu^2$ parameters} 
 In this section  we complete the previous analysis for   the cases where parameters $\mu_1^2$ and/or $\mu_2^2$ change sign.

 \subsection{Case $\mu_1^2< 0$ and $\mu_2^2>0$} 
 
Firstly, in this case the symmetric solution $\rho_1=\rho_2=0$  is a saddle point as it can be seen directly from (\ref{hessian}).

The solutions of the kind $\rho_1=0$ and $\rho_2\not=0$, namely,
$$\rho_2^2=\frac{2\mu_2^2}{\lambda_2}\,,$$
can be realized as minima if,
\begin{equation}
\gamma > \lambda_2\frac{\mu_1^2}{\mu_2^2}\,,
\label{muf}
\end{equation} 
as it can be seen from the expression of the hessian (\ref{hessian}). Obviously solutions of the form $\rho_2=0$ and $\rho_1\not=0$ can not be realized.

  Now let us analyze the case $\rho_1\not=0$ and $\rho_2\not=0$. In order to have non vanishing solutions we impose that expressions (\ref{solRho1}) and (\ref{solRho2}) to be positive, so that we arrive to the inequalities inside the cone, 
 \begin{equation}
 \gamma<\lambda_2\,\frac{\mu_1^2}{\mu_2^2}\,, \hspace{1cm}\gamma>\lambda_1\,\frac{\mu_2^2} {\mu_1^2}\,,\hspace{1cm} -\sqrt{\lambda_1\lambda_2}<\gamma<\sqrt{\lambda_1\lambda_2}\,, \label{ineqmu1} 
 \end{equation}
and this is  a non empty region of parameters, and furthermore, for the solutions belonging to this  set of inequalities we have a positive definite Hessian as it can be seen from (\ref{hessian0}). So in this region, the vacuum expectation values are given by (\ref{solRho1}) and (\ref{solRho2}),    Also we must look outside the cone $\gamma>\sqrt{\lambda_1\lambda_2}$, and we find the inequalities,
 \begin{equation}
 \gamma>\lambda_2\,\frac{\mu_1^2}{\mu_2^2}\,, \hspace{1cm}\gamma<\lambda_1\,\frac{\mu_2^2} {\mu_1^2}\,,\hspace{1cm} \gamma>\sqrt{\lambda_1\lambda_2}\,, \label{ineqmu2} 
 \end{equation}
 Clearly this set of inequalities are incompatible, and there is no solution of this kind outside the cone.

 Hence the phase space is simple in this case: we have  a phase of type IV inside the cone and below  the plane $\gamma=\lambda_2\frac{\mu_1^2}{\mu_2^2}$.

\subsection{Case $\mu_1^2>0$ and $\mu_2^2<0$}

This case  is entirely similar to the case above but swapping indices $1$ and $2$. the phase space  is drawn in Fig.~\ref{rightI}. Note that the phase IV has changed its location and phase III appears instead of phase III'.

 \subsection{Case $\mu_1^2,\mu_2^2<0$} 
 
 It is easy to see that in this case, if we assume that $\gamma^2<\lambda_1\lambda_2$, the region for $ \rho_1^2$ and $\rho_2^2$ being positive in (\ref{solRho1}) and (\ref{solRho2}) is that above both planes $ \gamma=\lambda_2\frac{\mu_1^2}{\mu_2^2}$ and $\gamma=\lambda_1\frac{\mu_2^2}{\mu_1^2}$, and hence there is no way to have the phase $I$ outside the cone. Furthermore, by setting any of the $\rho$'s to zero, the other can not reach a minimum. Hence the only possibility for the minimum is, \begin{equation} \rho_1=0\,,\hspace{3cm}\rho_2=0\,, \end{equation} which corresponds to the maximal symmetric vacuum. Similar situation occurs outside the cone, and only symmetric solution is possible.

\section{Instantons and Higgs Portal}

In this section we generalize the results given in III and IV for the Higgs portal (\ref{energyDensity}).   More specifically in phase I we have two minima separated by a potential barrier while in phases II and II' there are two asymmetric minima, one of them is a false vacuum. 

These two cases are of interest by the following reasons: 
\begin{itemize}
\item Potential  I is the classical Mexican hat potential  for which spontaneous symmetry breaking is restored by instantons effects. 
\item The potential II and II' is metastable and decays to a true vacuum. However, when it decays to true vacuum, it transfers information about the physical parameters to true vacuum. 
\end{itemize}

Interestingly the two cases mentioned above can be analyzed in the infrared limit assuming the boundary condition
\bb
\lim_{|x| \to \infty}  \varphi_{1,2} (x) \to \pm v_{1,2}. \label{xbc}
\ee

The equation  of motion are 
\eqb 
{\ddot \varphi_1} &=& -\mu_1^2 \varphi_1+ {\bar \lambda}_1 |\varphi_1|^2 \varphi_1 + {\bar \gamma} |\varphi_2|^2 \varphi_1, \label{y1}
\\
{\ddot \varphi_2} &=& -\mu_2^2  \varphi_2+ {\bar \lambda}_2 |\varphi_2|^2 \varphi_2 +{\bar  \gamma} |\varphi_1|^2 \varphi_2,  \label{y2}
\eqf
where 
\bb
{\bar \lambda}_{1,2} = \frac{\lambda_{1,2}}{\ell^2}, ~~~~~~~~~{\bar \gamma} =\frac{\gamma}{\ell^2}.
\ee

Although we cannot solve these equations analytically, we can try to find asymptotic solutions using the fact that both $\varphi_1$ and 
$\varphi_2 $ tend to $v_1$ and $v_2$. Using the condition (\ref{xbc}) and the fact that  
\begin{eqnarray}
v_1^2&=&\frac{-{\bar \lambda}_2\mu^2_1 +{\bar \gamma}\mu_2^2}{{\bar \gamma}^2 -{\bar \lambda}_1{\bar \lambda}_2}\,,\label{xsolRho1}\\
v_2^2&=&\frac{-{\bar \lambda}_1\mu^2_2 +{\bar \gamma}\mu_1^2}{{\bar \gamma}^2 -{\bar \lambda}_1{\bar \lambda}_2}\,, \label{xsolRho2}
\end{eqnarray}

The first and second equations become 
\eqb 
{\ddot \varphi_1} &=& -\left(\mu_1^2 -{\bar \gamma} v_2^2\right) \varphi_1+ {\bar \lambda}_1 |\varphi_1|^2 \varphi_1, \nonumber 
\\
{\ddot \varphi_2} &=& -\left(\mu_2^2 -{\bar \gamma} v_1^2\right) \varphi_2+ {\bar \lambda}_2 |\varphi_2|^2 \varphi_2.
\eqf
or in terms of the effective parameters 
\eqb 
{{\tilde \mu}_1}^2 &=& \left(\mu_1^2 -{\bar \gamma} v_2^2\right) = \frac{{\bar \lambda}_1({\bar \gamma} \mu_2^2 -{\bar \lambda}_2 \mu_1^2)}{{\bar \gamma}^2 -{\bar \lambda}_1 {\bar \lambda}_2} \nonumber 
\\
{\tilde \mu}_2^2 &=&  \left(\mu_1^2 -{\bar \gamma} v_2^2\right) = \frac{{\bar \lambda}_2({\bar \gamma} \mu_1^2 -{\bar \lambda}_1 \mu_2^2)}{{\bar \gamma}^2 -{\bar \lambda}_1 {\bar \lambda}_2}. \label{para12}
\eqf
Then the asymptotic solutions are 
\eqb 
{\bar \varphi}_ 1 (t) &=& \pm \sqrt{\frac{{\tilde \mu}_1^2}{{\bar \lambda}_1}} \tanh \left[  \sqrt{\frac{{\tilde \mu}_1^2 }{2}} (t-t_0) \right], 
\\
{\bar \varphi}_2 (t) &=& \pm \sqrt{\frac{{\tilde \mu}_2^2}{{\bar \lambda}_2}} \tanh \left[  \sqrt{\frac{{\tilde \mu}_1^2 }{2}} (t-t_0) \right], 
\eqf
which are two uncoupled (dilute)  instantons with effective parameters $({\tilde {\mu}}_1,{\bar \lambda}_1)$ and $({\tilde {\mu}}_2,{\bar \lambda}_2)$ respectively. 

The stability of the vacua in the Higgs portal must satisfy
\bb
{\tilde \mu^2}_1 >0, ~~~~~~{\tilde \mu^2}_2 >0. 
\ee
If  ${\tilde \mu^2}_1 = {\tilde \mu^2}_2>0$, then the spontaneous symmetry breaking is restored by tunneling of instantons and if 
${\tilde \mu^2}_1 \neq {\tilde \mu^2}_2>0$, the state of higher energy becomes unstable through barrier penetration and this is an example of a false vacuum \cite{coleman1}. 

As discussed in the previous sections, the physical implications that follow from our analysis are a consequence of the infrared approximation studied here and valid when (\ref{condicion1}) is fulfilled.

\section{Conclusions} 

The spontaneous symmetry breaking phenomenon is the cornerstone of the standard model, chiral symmetry breaking and so on. However, the connection between tunneling and the relationship of Yang-Mills theory is not fully understood and, for this reason in this paper we have introduced a reinterpretation of the  ultralocal Klauder limit that we call the infrared approximation in order to have a more physical view of the phenomenon. 

In this work we have studied the case in which the depth of the well is $\ell \sim \mu^{-1}$ in such a way that the excitations are either confined or undergo tunneling between one vacuum to another.

Tunneling by instantons is a phenomenon that appears only in $2D$ but from the point of view of the infrared approximation studied in this paper, instanton tunneling emerges naturally (see eq. (\ref{4x})). The relationship between these \lq \lq infrared instantons'' and Yang-Mills theory also appears quite naturally as shown in section II and follows from the gauge potential ansatz, namely $A^a_\mu = A^a_\mu (x^2)$, is $O(4)$ invariant producing the well-known dimensional reduction as in a central field in classical mechanics. 

Finally, a Higgs portal is considered as a way to understand the effect of scalar fields extras and the role that these can have in the case of cold dark matter. A discussion of this will be given elsewhere.

\section*{Acknowledgements}
 We would like to acknowledge the discussions and suggestions of J. L. Cort\'es, H. Falomir, F. A. Schaposnik and F. M\'endez. This work was partially supported by Dicyt 041831GR (J.G)., and by Fundacion ONCE with grant {\it Oportunidad al Talento} (J.L.S.).

\end{document}